\documentstyle[pre,aps,epsbox]{revtex}
\begin{document}
\draft
%
\twocolumn[\hsize\textwidth\columnwidth\hsize\csname@twocolumnfalse%
\endcsname
\title{
Spectral Properties of Three-Dimensional Quantum Billiards\\ 
with a Pointlike Scatterer
}
\author{
T. Shigehara
}
\address{
Computer Centre, University of Tokyo, Yayoi, 
Bunkyo-ku, Tokyo 113, Japan 
}
\author{
Taksu Cheon
}
\address{
Theory Division, Institute for Nuclear Study, 
University of Tokyo, 
Tanashi-shi, Tokyo 188 Japan
\\
and
\\
Laboratory of Physics, Kochi University of Technology,
Tosa Yamada, Kochi 782, Japan}
\date{February 7, 1997}
\maketitle
\begin{abstract}
We examine the spectral properties of three-dimensional 
quantum billiards with a single pointlike scatterer inside. 
It is found that the spectrum shows chaotic
(random-matrix-like) characteristics
when the inverse of the formal strength $\bar{v}^{-1}$ is 
within a band whose width increases parabolically
as a function of the energy.
This implies that the spectrum becomes 
random-matrix-like at very high energy irrespective to 
the value of the formal strength. 
The predictions are confirmed by numerical experiments 
with a rectangular box. 
The findings for a pointlike scatterer are applied to the 
case for a small but finite-size impurity. 
We clarify the proper procedure for its zero-size limit
which involves non-trivial divergence. 
The previously known results in one and two-dimensional 
quantum billiards with small impurities inside are also 
reviewed from the present perspective. 
\end{abstract}

\pacs{5.45.+b, 3.65.-w}
]
%
%
%

%
%
\section{Introduction}

The quantum billiard with pointlike scatterers 
inside is a solvable model which still retains
most of the interesting characteristics of 
non-integrable quantum physics.
The problem is based on obvious physical motivations. 
The billiard is a natural idealization
of the particle motion in bounded systems.
The one-electron problem in quantum dots is 
a possible setting which may be used as a single-electron memory,
a promising computational device in the future. 
It is now possible to actually construct such structures 
with extremely pure semiconductors thanks to the
rapid progress in the mesoscopic technology.
However, real systems are not free from impurities which 
affect the electron motion inside. 
In the presence of a small amount of contamination, 
even a single-electron problem in bounded regions 
becomes unmanageable.
The modeling of the impurities with pointlike scatterers is
expected to make the problem easy to handle without changing 
essential physics, at least at low energy. 
 
In spite of its seeming simplicity, the billiard problem 
with pointlike scatterers is known to possess
several non-trivial properties.
In two-dimensional billiards with a single 
pointlike scatterer, one observes phase reversion of 
wave function with the adiabatic rotation of the scatterer 
around certain points inside the billiard \cite{CS96a}. 
This can be regarded as the simplest manifestation 
of the geometrical phase or 
Berry phase \cite{B84,AA87}. 
Moreover, the two-dimensional quantum billiards 
with pointlike scatterers  possess the properties 
of ultra-violet divergences, scale anomaly and 
asymptotic freedom which are analogous to the
ones found in quantum field theories \cite{CS96b}. 
Also, there is a problem of so-called  
{\em wave chaos} \cite{S90,AS91};
Through its wave-like nature, the quantum particle can be diffracted
by pointlike scatterers, which should have no effect on the classical
motion of the particle \cite{RB81,ZK75,CC89}. 

A fundamental problem for quantum billiards with pointlike scatterers 
is to understand global behavior of the energy spectrum 
in the parameter space of particle energy $z$ and the strength 
of the scatterers $\bar v$. 
In particular, statistical properties of the spectrum are important  
because they reflect the degree of complexity (regularity or chaos) 
of underlying dynamics. For two dimension, the problem
has already been examined in details \cite{S94,SC96}.   
The chaotic spectrum (level statistics of random-matrix 
theory \cite{DM63,M67,BGS84}) appears along the ``logarithmic strip''
in the parameter space $(z, \bar v^{-1})$. 
More precisely, 
the effects of a pointlike scatterer with formal strength $\bar{v}$ 
are most strongly observed in the eigenstates 
with an eigenvalue $z$ such that  
\begin{eqnarray}
\label{1}
\frac{M}{2\pi} \ln \frac{z}{\Lambda} \simeq \bar{v}^{-1}, 
\end{eqnarray}
where $M$ is the mass of a particle moving in the billiard and  
$\Lambda$ is an arbitrary mass scale. 
Eq. (\ref{1}) indicates 
that the maximal physical coupling is attained at the value of
formal coupling that varies with the logarithmic dependence of the 
particle energy.
This energy-dependence is a manifestation of a phenomenon known 
as the scale anomaly, 
or the quantum mechanical breaking of 
scale invariance \cite{TH79,J95}: 
In two dimension, the physics is expected to be energy-independent, 
since the kinetic term (Laplacian) and the zero-range 
interaction (a $\delta$-potential) are scaled 
in the same manner under a transformation of length scale. 
However, the quantization breaks a scale invariance, and 
as a result, the strong coupling region shifts 
with a logarithmic dependence of energy. 
The condition (\ref{1}) also shows that, for any value of formal 
strength $\bar{v}$, the system approaches to the empty billiard
when the energy increases. Thus the system possesses the property 
of the asymptotic freedom.  

Quantum-mechanical billiard problems with pointlike scatterers 
inside can be defined for spacial dimension $d \leq 3$. 
Contrary to the two-dimensional case, spectral properties 
in three dimension have been scarcely studied so far. 
The main purpose of this paper is precisely to fill this void.
The logarithmic energy-dependence of the strong coupling 
region observed in two dimension has its origin in the energy
independence of the average level density of the system.
Since the level density is proportional to the square root of
the energy in three dimension, one expects
substantially different spectral properties.
In this paper, we find that this is indeed the case. 
It is shown that the value of formal strength 
which induces the maximal coupling is independent of the 
particle energy,
whereas the width of the strip on which the strong coupling 
is attained broadens with square-root dependence of energy.  
This means that, in three dimension, for any $\bar{v}$ ($\neq 0$), 
the system exhibits chaotic spectra at the high energy limit. 

Another objective of this paper is to relate the findings in
the purely pointlike scatterers to the realistic situation of
small but finite-size impurities. 
For the pointlike scatterers, 
the condition for the strong coupling also depends on the mass 
scale $\Lambda$ which is introduced in the process of regularization. 
This reflects the fact that formal strength $\bar{v}$ 
does not have a direct relation to the observables as it stands. 
In order to clarify the physical meaning of $\bar{v}$, 
we begin by approximating a finite-range potential 
with a $\delta$-potential within a truncated basis. 
The size of the truncation depends on the range of potential. 
We then obtain a relation between the formal and bare strengths, 
the latter of which corresponds to the strength of 
the $\delta$-potential within the truncation. 
The relation enables us to apply the results for  
pointlike scatterers to finite-range cases. 
Moreover, it clarifies the proper procedure and physical 
meaning of the zero-size limit of the finite-size potential 
in an intuitive fashion.

The paper is organized as follows. 
In Sec. II, we deduce, from a general perspective
without any assumption on the shape of billiards, 
the strong coupling condition in three-dimensional billiards
under which the effect of a pointlike scatterer 
becomes substantial.   
In Sec. III, we consider the case for a small but 
finite-size scatterer. 
By examining a relation between the formal strength of the scatterer 
and the energy eigenstates of finite-size potential, we rewrite 
the condition for the strong coupling in terms of the observables. 
The previously known results for one and two dimensions are reviewed 
from the present point of view.
We clarify the proper procedure and meaning 
of the zero-size limit of
finite-size potential in one, two and three dimensions.
We test the predictions in Sec. II with the numerical 
calculations in Sec. IV. 
We look at the level statistics of rectangular box 
with a single pointlike scatterer inside.  
In particular, the case where the scatterer is 
located at the center of the box is examined in details. 
We summarize the present work in Sec.V. 

%
%
\section{Condition for Strong Coupling in terms of Formal Strength}

Consider a quantum particle of mass $M$ moving  
in a three-dimensional billiard of volume $V$. 
The eigenvalues and eigenfunctions of this system 
are denoted by 
$E_{n}$ and $\varphi_{n}(\vec{x})$; 
\begin{equation}
\label{2}
-\frac{\nabla^{2}}{2M}\varphi_{n}(\vec{x}) = E_{n} \varphi_{n}(\vec{x}) 
\ \ \ \ \ (n=1,2,3,\cdots).
\end{equation}
We impose the Dirichlet boundary condition to the wave functions
$\varphi_{n}$ at the billiard surface.
The average level density at energy $z$ has square-root energy 
dependence; 
\begin{equation}
\label{3}
\rho_{av}(z)=\frac{M^{3/2}V}
{2^{1/2}\pi^{2}}\sqrt{z}. 
\end{equation}
Suppose that a single pointlike scatterer 
is placed at $\vec{x}=\vec{x}_{0}$ inside the billiard. 
Despite the simplicity of a contact interaction, 
the Schr\"{o}dinger equation suffers from 
short-distance singularities at the location of the 
scatterer, which needs to be renormalized. 
This can be done in most mathematically satisfying fashion 
in the framework of the self-adjoint extension theory
of a symmetric operator in functional analysis. 
Details are given elsewhere (see Ref.\cite{SC96}).  
We just present the relevant results.
Starting with the formulation of Zorbas \cite{Z80}, 
we obtain the equation for the eigenvalues of the system,
$z_{n}$ $(n=1,2,3,\cdots)$, as 
\begin{eqnarray}
\label{4}
\bar{G}(z) = \bar{v}^{-1}, 
\end{eqnarray}
where 
\begin{eqnarray}
\label{5}
\bar{G}(z) \equiv \sum_{n=1}^{\infty} \varphi_{n}(\vec{x}_{0})^{2} 
\left( \frac{1}{z-E_{n}}+\frac{E_{n}}{E_{n}^{2}+\Lambda^2} \right). 
\end{eqnarray}
In Eq. (\ref{5}), $\bar{v}$ is the formal strength 
of the pointlike scatterer and $\Lambda$ is an arbitrary 
mass scale that arises in the renormalization. 
The formal strength $\bar{v}$ does not have a direct 
relation to physical observables as it stands. 
Its relation to physical strength of 
the scatterer is discussed later in Sec. III.  
Here we just mention the following 
two points; 
(1) To ensure the self-adjointness of the Hamiltonian 
for the system defined by Eq. (\ref{4}), one has to take 
$\bar{v}$ to be independent of the energy, 
and (2) In the limit of $\bar{v} \rightarrow 0$, 
the system approaches the empty billiard. 

The second term of $\bar{G}(z)$ in Eq. (\ref{5}) is 
independent of the energy $z$. 
It plays an essential role in making the problem well-defined; 
The infinite series in Eq. (\ref{5}) does not converge without 
the second term.
For spacial dimension $d\geq 4$, 
the summation in Eq. (\ref{5}) diverges. 
This reflects the fact that 
the billiard problem with pointlike scatterers  
is not well-defined for $d\geq 4$ in quantum mechanics.
Within any interval between two neighboring unperturbed eigenvalues,
$\bar{G}(z)$ is a monotonically decreasing function 
that ranges over the whole real number.  Therefore,
Eq. (\ref{4}) has a single solution on each interval. 
The eigenfunction corresponding to an eigenvalue $z_{n}$ is written 
in terms of the Green's function of the empty billiard as
\begin{eqnarray} 
\label{6} 
\psi_{n}(\vec{x}) & \propto &
G^{(0)}(\vec{x},\vec{x}_{0};z_{n}) \nonumber \\
& = & \sum_{k=1}^{\infty} 
\frac{\varphi_{k}(\vec{x_{0}})}
     {z_{n}-E_{k}}\varphi_{k}(\vec{x}). 
\end{eqnarray}
This shows that if a perturbed eigenvalue $z_n$ is close 
to an unperturbed one $E_n$ (or $E_{n+1}$), 
then the corresponding eigenfunction $\psi_n$ 
is not substantially different from 
$\varphi_n$ (or $\varphi_{n+1}$).
Thus, the disturbance by a pointlike scatterer is  
restricted to eigenstates with an eigenvalue around which 
$\bar{G}(z)$ has an inflection point.  
This is because each inflection point of $\bar{G}(z)$ is expected 
to appear, on average, around the midpoint on the interval between 
two neighboring unperturbed eigenvalues. 
Let $(\tilde{z},\bar{G}(\tilde{z}) )$ be one of 
such inflection points of $\bar{G}(z)$; 
$\ \ \tilde{z} \simeq (E_{m}+E_{m+1})/2 \ $ for some $m$.  
In this case, 
the contributions on $\bar{G}(\tilde{z})$ from 
the terms with $n \simeq m$ cancel each other, and 
we can replace the summation in Eq. (\ref{5}) by a principal 
integral with a high degree of accuracy;   
\begin{eqnarray}
\label{7}
\bar{G}(\tilde{z}) &\simeq& \bar{g}(\tilde{z}) 
\\ \nonumber
&\equiv& 
\langle \varphi_{n}(\vec{x}_{0})^{2} \rangle 
P  \int_{0}^{\infty} 
\left( \frac{1}{\tilde{z}-E}+\frac{E}{E^{2}+\Lambda^2} \right)
\rho_{av}(E)dE, 
\end{eqnarray} 
where we have defined a continuous function $\bar{g}(z)$ 
which behaves like an interpolation of the inflection 
points of $\bar{G}(z)$. In Eq. (\ref{7}), 
$\langle \varphi_{n}(\vec{x}_{0})^{2} \rangle$ is 
the average value of $\varphi_{n}(\vec{x}_{0})^{2}$ among 
various $n$. For a generic position of the scatterer, one has  
$\langle \varphi_{n}(\vec{x}_{0})^{2} \rangle \simeq 1/V$. 
Notice that $\bar{G}(z) \simeq \bar{g}(z)$ is valid only around 
the inflection points of $\bar{G}(z)$. 
Using an elementary indefinite integral 
\begin{eqnarray}
\label{8}
& &
\int \left( \frac{1}{z-E}+\frac{E}{E^{2}+\Lambda^2} \right) \sqrt{E} dE
\\ \nonumber
 & = & 
\sqrt{z} \ln \left| \frac{\sqrt{z}+\sqrt{E}}{\sqrt{z}-\sqrt{E}} \right| 
-  \frac{1}{2}\sqrt{\frac{\Lambda}{2}} 
\ln \left( \frac{E+\sqrt{2\Lambda E}+\Lambda}{E-\sqrt{2\Lambda E}+\Lambda} 
\right) 
\\ \nonumber
& - & 
\sqrt{\frac{\Lambda}{2}}
\left\{ \arctan(\sqrt{\frac{2E}{\Lambda}}+1)+
\arctan(\sqrt{\frac{2E}{\Lambda}}-1) \right\} 
\end{eqnarray}
for $z>0$, we obtain 
\begin{eqnarray}
\label{9}
\bar{G}(\tilde{z}) \simeq -\frac{M^{3/2}\Lambda^{1/2}}{2\pi}. 
\end{eqnarray} 
The first term in Eq. (\ref{8}), which depends on the energy $z$,  
disappears both at $E=0$ and $E=\infty$. 
As a result, the average value of $\bar{G}(z)$ 
at the inflection points is independent of the energy. 
Eq. (\ref{9}) indicates that the maximal coupling of 
a pointlike scatterer is attained with the formal strength 
${\bar v}$ which satisfies 
\begin{eqnarray}
\label{10}
\bar{v}^{-1} \simeq - \frac{M^{3/2}\Lambda^{1/2}}{2\pi}.   
\end{eqnarray} 

\begin{figure}
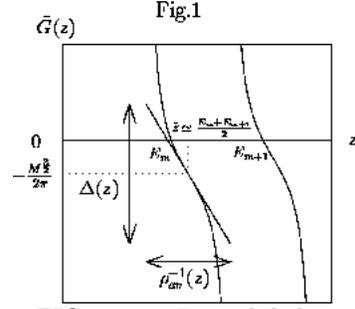

\center\psbox[hscale=0.5,vscale=0.5]{dmfig1.epsf}
\caption{
Typical behavior of $\bar{G}(z)$ with mass scale 
$\Lambda=1$ in Eq. (5) and 
its linearized version is shown as a function of $z$. 
The latter is obtained by expanding $\bar{G}(z)$ at 
its inflection point $\tilde{z}$ 
on the interval $E_m<z<E_{m+1}$. 
The coordinate of the inflection point is given by 
$\left( \tilde{z},\bar{G}(\tilde{z}) \right) \simeq  
( (E_m+E_{m+1})/2,-M^{\frac{3}{2}}/2\pi )$. 
Strong coupling is attained when 
$\bar{v}^{-1}$ takes a value within 
the range of the linearized function. 
}
\label{fig1}
\end{figure}
The ``width'' of the strong coupling region can be estimated 
by considering a linearized eigenvalue equation. 
Expanding $\bar{G}(z)$ around $\tilde{z}$, 
we can rewrite Eq. (\ref{4}) as 
\begin{eqnarray}
\label{11}
\bar{G}(\tilde{z})+
\bar{G}'(\tilde{z})(z-\tilde{z}) \simeq \bar{v}^{-1}  
\end{eqnarray} 
or
\begin{eqnarray}
\label{12}
\bar{G}'(\tilde{z})(z-\tilde{z}) \simeq \bar{v}^{-1}  +
\frac{M^{3/2}\Lambda^{1/2}}{2\pi}.   
\end{eqnarray} 
In order to ensure that the perturbed eigenvalue $z_m$ is 
close to $\tilde{z}$, the range of 
$\bar{v}^{-1} + M^{3/2}\Lambda^{1/2}/2\pi$ has to be restricted to 
\begin{eqnarray}
\label{13}
\left| \bar{v}^{-1} + \frac{M^{3/2}\Lambda^{1/2}}{2\pi} \right| 
\alt \frac{\Delta(\tilde{z})}{2}
\end{eqnarray} 
where the width $\Delta$ is defined by    
\begin{eqnarray}
\label{14}
\Delta(\tilde{z}) \equiv 
\left| \bar{G}'(\tilde{z}) \right| \rho_{av}(\tilde{z})^{-1}. 
\end{eqnarray} 
This is nothing but the variance of the linearized $\bar{G}(z)$ 
on the interval between the two 
unperturbed eigenvalues just below and above $\tilde{z}$ 
(see Fig. \ref{fig1}).
The width can be estimated by the average level density 
at the energy under consideration as follows;  
\begin{eqnarray}
\label{15}
\Delta(\tilde{z}) & = & \sum_{n=1}^{\infty} 
\frac{\varphi_{n}(\vec{x}_{0})^{2}}{(\tilde{z}-E_{n})^2} 
\rho_{av}(\tilde{z})^{-1} \nonumber \\
& \simeq &
\langle \varphi_{n}(\vec{x}_{0})^{2} \rangle 
\sum_{n=1}^{\infty} 
\frac{2\rho_{av}(\tilde{z})^{-1}}
{\{(n-\frac{1}{2})\rho_{av}(\tilde{z})^{-1}\}^2} \nonumber \\
& = & 
\pi^{2} \langle \varphi_{n}(\vec{x}_{0})^{2} \rangle \rho_{av}(\tilde{z}) 
\nonumber \\
& \simeq &
\frac{M^{3/2}}{2^{1/2}}\sqrt{\tilde{z}}.
\end{eqnarray}
We have implicitly assumed in Eq. (\ref{15}) that 
the unperturbed eigenvalues are distributed with a mean 
interval $\rho_{av}(\tilde{z})^{-1}$ in the whole energy 
region. This assumption is quite satisfactory, 
since the denominator of $\bar{G}'(z)$ is 
of the order of $(z-E_n)^2$, indicating that the summation 
in Eq. (\ref{15}) converges rapidly. 

\begin{figure}
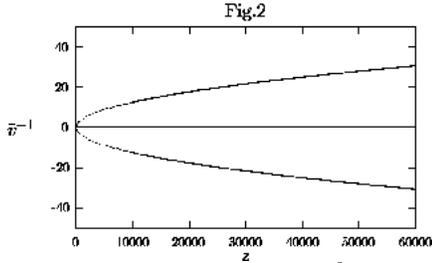

\center\psbox[hscale=0.5,vscale=0.5]{dmfig2.epsf}
\caption{
Plot of $\bar{v}^{-1}=-M^{\frac{3}{2}}/2\pi \pm \Delta (z)/2$. 
The effects of a pointlike scatterer on the energy spectrum 
are expected to appear in the eigenstates in the region 
between both curves. 
}
\label{fig2}
\end{figure}
We recognize from Eqs. (\ref{13}) and (\ref{15}) that 
the effects of a pointlike scatterer of formal strength $\bar{v}$ 
are substantial only in the eigenstates with eigenvalues $z$ 
such that 
\begin{eqnarray}
\label{16}
\left| \bar{v}^{-1} + \frac{M^{3/2}\Lambda^{1/2}}{2\pi} \right| 
& \alt & \frac{\Delta(z)}{2} \nonumber \\
& \simeq &  
\frac{M^{3/2}}{2^{3/2}}\sqrt{z}.
\end{eqnarray}
The width $\Delta$ is proportional to the average level density, and 
as a result, it broadens with square-root energy dependence 
(see Fig. \ref{fig2}).  
This can be understood from another perspective, 
by considering a scale transformation of 
a heuristic Hamiltonian with a $\delta$-potential; 
\begin{eqnarray}
\label{17}
H = -\frac{\nabla^2}{2M}+v \delta (\vec{x}-\vec{x}_0). 
\end{eqnarray}
Although the Hamiltonian (\ref{17}) is not well-defined 
in case of spacial dimension $d \geq 2$, 
we proceed further for the moment and return to this point 
in Sec. III. 
Under a scale transformation $\vec{x}\longrightarrow \vec{x}/a$, 
the Hamiltonian (\ref{17}) is transformed to 
\begin{eqnarray}
\label{18}
H \longrightarrow a^2 
\left(
  -\frac{\nabla^2}{2M}+ a v \delta (\vec{x}-\vec{x}_0) 
\right). 
\end{eqnarray}
Since the energy $z$ scales as $z \longrightarrow a^2 z$, 
the strength $v$ which scales as  $v \longrightarrow a v$ must
have square-root energy dependence, which explains Eq. (\ref{16}). 

The findings in this section are summarized as follows; 

(1) For a three-dimensional billiard, 
the effect of a pointlike scatterer on spectral properties is 
maximal when the formal strength of the scatterer satisfies 
$\bar{v}^{-1} \simeq -M^{3/2}\Lambda^{1/2}/2\pi$, 
irrespective to the energy $z$. 

(2) The width $\Delta$ 
(or an allowable error in $\bar{v}^{-1}$ to look for  
the effect) increases with square-root energy dependence.  

 From these two, we conclude: 

(3) For any value of formal strength $(\bar{v}\neq 0)$, 
a pointlike scatterer tends to disturb a particle 
motion in billiards, as the particle energy increases;  
The physical strength increases proportional to the square-root 
of the energy.
This makes a sharp contrast to the asymptotic freedom 
observed in two-dimensional billiards. 

Before closing this section, we give a few words 
on the shape of the billiard. 
Our implicit assumption for the shape is that 
the average level density of the empty billiard  
is dominated by the volume term, which has 
a square-root dependence on energy. 
The assumption is justified for a generic 
three-dimensional billiard which has 
the same order of length scale in each direction,      
irrespective to a full detail of the shape of the billiard.  
If one length scale is substantially smaller than the other two, 
the surface term dominates the average level density 
in the low energy region. 
As a result, the spectral property at low energy 
is expected to change with the logarithmic energy-dependence  
which is specific to two dimension [see Eq. (\ref{1})]. 

%
%
\section{Formal, Bare and Effective Strengths}

As stated in the previous section, 
$\bar{v}$ in Eq. (\ref{4}) does not have a 
direct relation to physical observables as it stands. 
The main purpose of this section is to clarify 
the physical meaning of the formal strength $\bar{v}$. 
For this end, we relate the formal strength to 
a strength defined through a more realistic 
potential with small but finite range. 
The relation makes it possible to apply the findings 
in the previous section to the finite-size impurities. 
The previously published results in two dimension \cite{SC96} 
and the well-known elementary results in one dimension
are also reviewed from the present perspective. 

We first point out that the definition of 
the formal strength is not unique. 
Indeed, Eq. (\ref{5}) is not 
a unique candidate for $\bar{G}(z)$; 
It can be defined by any convergent series  
for $z \neq E_n$ which has a form, 
\begin{eqnarray}
\label{19}
\bar{G}(z) = \lim_{N\longrightarrow\infty} \sum_{n=1}^{N} 
\left( \frac{\varphi_{n}(\vec{x}_{0})^{2}}{z-E_{n}}+ 
f_n \right). 
\end{eqnarray}
Here, $f_n$ is an arbitrary quantity 
independent of the energy $z$, whereas it may 
depends, in general, on $E_n$ and $\varphi_{n}(\vec{x}_{0})$.
The first term in the parenthesis on RHS in Eq. (\ref{19}) 
does not converge as $N\rightarrow\infty$
in spacial dimension $d=2$, $3$. 
This means that $f_n$ should be taken as a counter-term which 
cancels the divergence of the first term. 
Once such a series $\{ f_n \}$ is chosen, 
one can define an equation, $\bar{G}(z)=\gamma$, with an energy-independent 
constant $\gamma$. This gives a possible eigenvalue equation 
for the billiard with a pointlike scatterer of a certain fixed 
(energy-independent) coupling strength. 
It is obvious that, even with another choice of series, 
say $\{ \tilde{f}_n \}$, 
the same eigenvalue equation can be reproduced 
by shifting the value of $\gamma$ by   
$\displaystyle \sum_{n=1}^{\infty}(\tilde{f}_n-f_n)$. 
One possible choice of $f_n$ is given by 
\begin{eqnarray}
\label{20}
f_n = \varphi_{n}(\vec{x}_{0})^{2}\frac{E_n}{E_n^2+\Lambda^2} 
\end{eqnarray}
with an arbitrary real number $\Lambda$ ($\neq 0$). 
This choice along with the definition $\bar{v} \equiv 1/\gamma$
gives precisely the original eigenvalue problem Eqs. (\ref{4}) 
and (\ref{5}).
Clearly, this $\bar{v}$ is a mathematical 
quantity whose physical interpretation is not immediately evident. 

To reveal the meaning of 
the formal strength $\bar{v}$ in Eq. (\ref{4}), 
we begin by approximating low-energy spectra     
(eigenvalues and eigenfunctions) 
of a finite-range potential by that of a zero-range interaction. 
Suppose that a small but finite-size scatterer of volume $\Omega$ is 
located at $\vec{x}=\vec{x}_0$ inside a three-dimensional billiard 
of volume $V$. 
We describe the scatterer in terms of 
a potential which has a constant strength on
a region $\Omega$; 
\begin{equation}
\label{21}
U(\vec{x}) = 
\left \{
\begin{array}{ll}
U_0, & \ \ \ \vec{x} \in \Omega \\
0  , & \ \ \ \vec{x} \in  V-\Omega, 
\end{array}
\right.
\end{equation}
where the regions of the potential and the outer billiard 
are denoted by the same symbols as the volumes. 
We assume that 
the scatterer has the same order of size, say $R$, in each 
spacial direction, and also assume 
that the volume of the scatterer is substantially smaller 
than that of the outer billiard; 
$\Omega \simeq R^3 \ll V$. 
In this case, 
the scatterer behaves as pointlike at low energy $z\ll E_{N(\Omega)}$,  
where $E_{N(\Omega)}$ is estimated as 
\begin{eqnarray}
\label{22}
E_{N(\Omega)} & \simeq & \frac{1}{MR^2} \nonumber \\
& \simeq & \frac{1}{M\Omega^{2/3}}. 
\end{eqnarray}
Furthermore, the coupling of higher energy states 
than $E_{N(\Omega)}$ to the low-energy states 
is weak, since wave functions with wavelength shorter than 
$R$ oscillate within the scatterer. 
This means that the low-energy states ($z\ll E_{N(\Omega)}$)  
can be described by the Hamiltonian in terms of  
a $\delta$-potential, Eq. (\ref{17}),   
with the coupling strength 
\begin{equation}
\label{23}
v \equiv U_0  \Omega,  
\end{equation}
together with {\em a basis truncated at} $E_{N(\Omega)}$. 
The truncation of basis is crucial for the present argument. 
In case of spacial dimension $d \geq 2$,  
a $\delta$-potential 
is not well-defined in the full unperturbed basis. 
This is clear from the fact that the summation in Eq. (\ref{24}) 
diverges in the limit of $\Omega\longrightarrow 0$ (hence as 
$N(\Omega)\longrightarrow \infty$). 
The finiteness of the scatterer introduces 
an ultra-violet cut-off in a natural manner, and as a result,  
the low-energy spectra can be reproduced by the Hamiltonian (\ref{17}) 
within a suitably truncated basis. 

In an analogy to the terminology of the field theories,
we call the coupling $v$ as the {\em bare strength},
since it appears as the coefficient
of the $\delta$-potential within a given truncation \cite{NT00}.
The bare strength $v$ can be related to 
formal strength $\bar{v}$ as follows. 
Within the truncated basis \{ $\varphi_n (\vec{x})$ ; 
$n = 1, 2, ..., N(\Omega)$\}, 
the eigenvalues of the Hamiltonian (\ref{17}) are determined by 
\begin{equation}
\label{24}
\sum_{n=1}^{N(\Omega)} 
\frac{\varphi_{n}(\vec{x}_0)^2}{z-E_n} = v^{-1}.
\end{equation}
Inserting Eq. (\ref{24}) into Eq. (\ref{4}) with Eq. (\ref{5}), 
we obtain 
\begin{eqnarray}
\label{25}
\bar{v}^{-1} & = & v^{-1} +   
\sum_{n=1}^{N(\Omega)} 
\varphi_{n}(\vec{x}_{0})^{2}\frac{E_{n}}{E_{n}^2+\Lambda^2} \nonumber \\
&& + \sum_{n=N(\Omega)+1}^{\infty}
\varphi_{n}(\vec{x}_{0})^{2}\left(
\frac{1}{z-E_{n}}+\frac{E_{n}}{E_{n}^2+\Lambda^2}\right).   
\end{eqnarray}
Eq. (\ref{25}) gives an exact relation between formal and
bare strengths.    
In order to have further insight on Eq. (\ref{25}), 
we take an average for $\varphi_{n}(\vec{x}_{0})^{2}$ 
among various $n$, $\langle \varphi_n (\vec{x}_0)^2 \rangle \simeq 1/V$, 
and replace the remaining summations on RHS by integrals. 
We then have  
\begin{eqnarray}
\label{26}
\bar{v}^{-1} & \simeq & 
v^{-1} +   \langle \varphi_n (\vec{x}_0)^2 \rangle 
\left\{ \int_{0}^{E_{N(\Omega)}} \frac{E}{E^2+\Lambda^2}\rho_{av}(E)dE 
\right.
\nonumber \\
&& +  \left.
\int_{E_{N(\Omega)}}^{\infty} 
\left(\frac{1}{z-E}+\frac{E}{E^2+\Lambda^2}\right)\rho_{av}(E)dE \right\}. 
\end{eqnarray}
Using Eq. (\ref{3}), along with elementary integrals 
\begin{eqnarray}
\label{27}
F_1^{(3)}(z,E) 
& \equiv & \int \frac{\sqrt{E}}{z-E} dE \nonumber \\
& = & 
\sqrt{z} \ln \left| \frac{\sqrt{z}+\sqrt{E}}{\sqrt{z}-\sqrt{E}} \right| 
- 2\sqrt{E}, \ \ \ \ (z>0),  \\
\label{28}
F_2^{(3)}(E) 
& \equiv & \int \frac{E}{E^{2}+\Lambda^2} \sqrt{E} dE \nonumber \\
=
&2\sqrt{E}& -  
\frac{1}{2}\sqrt{\frac{\Lambda}{2}} 
\ln \left( \frac{E+\sqrt{2\Lambda E}+\Lambda}{E-\sqrt{2\Lambda E}+\Lambda} \right) 
\nonumber \\
-&\sqrt{\frac{\Lambda}{2}}&
\left\{ \tan^{-1}(\sqrt{\frac{2E}{\Lambda}}+1)+
\tan^{-1}(\sqrt{\frac{2E}{\Lambda}}-1) \right\}, 
\end{eqnarray}
we can rewrite Eq. (\ref{26})  as 
\begin{eqnarray}
\label{29}
\bar{v}^{-1} & \simeq & 
v^{-1} - \frac{M^{3/2}\Lambda^{1/2}}{2\pi}-\frac{M^{3/2}}{2^{1/2}\pi^2}
F_1^{(3)}(z,E_{N(\Omega)}). 
\end{eqnarray}
In Eq. (\ref{27}), the first term on RHS is negligible 
in case of $z \ll E$. Hence, at low energy 
$z \ll E_{N(\Omega)}$, we have 
\begin{eqnarray}
\label{30}
\bar{v}^{-1} & \simeq & v^{-1} 
- \frac{M^{3/2}\Lambda^{1/2}}{2\pi}+
\frac{2^{1/2}M^{3/2}}{\pi^2}\sqrt{E_{N(\Omega)}}. 
\end{eqnarray}
The third term on RHS in Eq. (\ref{30}) diverges as $E_{N(\Omega)}$ increases. 
This is exactly the same divergence which we observe in the summation 
in Eq. (\ref{24}) [or Eq. (\ref{19})] 
with opposite sign. This ensures the convergence 
of $\bar{G}(z)$ in Eq. (\ref{5}). 
Using Eq. (\ref{22}), we arrive at
\begin{eqnarray}
\label{31}
\bar{v}^{-1} & \simeq & v^{-1} 
- \frac{M^{3/2}\Lambda^{1/2}}{2\pi}+
\frac{2^{1/2}M}{\pi^2\Omega^{1/3}}.
\end{eqnarray}

In order to reproduce a zero-range scatterer 
with a fixed formal strength $\bar{v}$ ($\neq 0$), 
the RHS of Eq. (\ref{31}) needs to converge as $\Omega$ shrinks.  
This means that, for small $\Omega$, 
$v$ should take a form, 
\begin{eqnarray}
\label{32}  
v(\Omega) = 1/\left( - \frac{C}{\Omega^{1/3}} + r(\Omega) \right). 
\end{eqnarray}
The first term in the denominator is a counter term 
that cancels the divergence of the third term on RHS in Eq. (\ref{31}); 
\begin{eqnarray}
\label{33}
C \simeq \frac{2^{1/2}M}{\pi^2}.  
\end{eqnarray}
[More precisely, $C$ should be taken to cancel the divergence 
which appears in the summation in Eq. (\ref{24}).] 
The remnant quantity $r(\Omega)$ 
in the denominator is a regular function 
which converges as $\Omega\longrightarrow 0$.  
In the zero-size limit, 
the finite-size scatterer shrinks into a pointlike one with 
formal strength  
\begin{equation}
\label{34}
\bar{v}^{-1} \simeq r(0) - \frac{M^{3/2}\Lambda^{1/2}}{2\pi}.   
\end{equation}
In terms of the potential height $U_0$, 
Eq. (\ref{32}) is rewritten as 
\begin{eqnarray}
\label{35}  
U_0 (\Omega) & = & 1/\left( - C \Omega^{2/3} + r(\Omega) \Omega \right)  
\nonumber \\
& = & 1/\left( - C \Omega^{2/3} + r(0) \Omega + o(\Omega) \right). 
\end{eqnarray}
Since $C$ is positive, we obtain
\begin{eqnarray}
\label{36}
\left\{ 
\begin{array}{lll}
v(\Omega)    & \longrightarrow & - 0, \\
U_0 (\Omega) & \longrightarrow & - \infty,  
\end{array}
\right.
\quad \mbox{as $\Omega \longrightarrow 0$.}
\end{eqnarray}
Eq. (\ref{36}) indicates that 
the potential has to be negative in the zero-size limit, 
irrespective to a form of $r(\Omega)$.  
This is consistent with the fact that 
a pointlike scatterer with an arbitrary formal strength 
$\bar{v}$ ($\neq 0$) 
sustains a single eigenstate with an eigenvalue smaller than $E_1$. 
A seemingly plausible limit, $\Omega\longrightarrow 0$ along with 
keeping $U_0 \Omega$ constant, 
is not allowable in the case of three dimension; 
Such a limit induces too strong a potential 
to define a quantum mechanical Hamiltonian for a pointlike scatterer. 
Notice that Eq. (\ref{36}) does not exclude a possibility 
of strong repulsion $U_0 \gg 0$ on a small-size region $\Omega \neq 0$.  
Indeed, Eq. (\ref{32}) [or Eq. (\ref{35})] does not impose any restriction 
on $v$ (or $U$) for any {\em finite} $\Omega$. 
As long as $\Omega$ is finite, 
one can reproduce even a strong repulsion by taking  
$r(\Omega)$ as slightly larger than $C/\Omega^{1/3}$. 
Such $r(\Omega)$ is, in general, a very large positive quantity which  
diverges to $+\infty$ when $\Omega$ shrinks into the zero-size
together with positively fixed $U_0$.   

Combining the findings in the current and previous sections, 
we can deduce the condition for the strong coupling for
a finite-size scatterer. 
Inserting Eq. (\ref{30}) or Eq. (\ref{31}) into Eq. (\ref{16}), 
we obtain 
\begin{eqnarray}
\label{37}
\left| 
v^{-1} + 
\frac{2^{1/2}M^{3/2}}{\pi^2}
\sqrt{E_{N(\Omega)}} 
\right| 
\alt 
\frac{\Delta(z)}{2} 
\end{eqnarray}
or 
\begin{eqnarray}
\label{38}
\left| v^{-1} +
\frac{C}{\Omega^{1/3}}
\right| 
\alt 
\frac{\Delta(z)}{2}  
\end{eqnarray}
for $z\ll E_{N(\Omega)}$.
Eq. (\ref{38}) indicates that $v^{-1} \simeq -C/\Omega^{1/3}$ is 
the condition for the strong coupling, and hence that 
the effects of a finite-size scatterer at low energy 
most strongly appear when it is weakly attractive, 
namely, when the bare strength $v$ is slightly negative.  
In the zero-size limit, the condition (\ref{38}) is equivalent to 
\begin{eqnarray}
\label{39}
\left| r(0) \right| \alt \frac{\Delta(z)}{2}.  
\end{eqnarray}
Eq. (\ref{39}) shows that it is the inverse of $r(0)$ that represents 
a direct measure of coupling strength of the zero-size limit of a scatterer. 
This naturally leads us to a definition of the {\em effective strength} 
of a pointlike scatterer by 
\begin{eqnarray}
\label{40}
v_{e\!f\!f}\equiv 1/r(0). 
\end{eqnarray}
Using the effective strength $v_{e\!f\!f}$, 
we can rewrite Eq. (\ref{39}) as 
\begin{eqnarray}
\label{41}
\left| v_{e\!f\!f}^{-1} \right| \alt \frac{\Delta(z)}{2}. 
\end{eqnarray}
It can be observed from Eq. (\ref{34}) that,   
if the origin of $\bar{v}^{-1}$-axis is shifted 
to the strong coupling value $-M^{3/2}\Lambda^{1/2}/2\pi$, 
the formal strength 
$\bar{v}$ is identical to $v_{e\!f\!f}$. 
We can also say that $v_{e\!f\!f}^{-1}$ is   
a ``distance'' to the strong coupling value $v^{-1} \simeq -C/\Omega^{1/3}$, 
which is, in general, a large negative quantity for small $\Omega$. 
Inserting Eqs. (\ref{15}), (\ref{23}) and (\ref{33}) into Eq. (\ref{38}), 
we have  
\begin{eqnarray}
\label{42}
\left| 
\frac{1}{U_0\Omega} + \frac{2^{1/2}M}{\pi^2\Omega^{1/3} }
\right| 
\alt 
\frac{M^{3/2}}{2^{3/2}}\sqrt{z}. 
\end{eqnarray}
Eq. (\ref{42}) is the condition for the strong coupling 
in terms of the ``observables'';   
At low energy where a finite-size scatterer can be approximated
by a pointlike one  
($z \ll 1/M\Omega^{2/3}$), 
a particle of mass $M$ moving in three-dimensional billiards 
is most strongly coupled to a finite-size ($\simeq \Omega$) scatterer 
of potential height $U_0$ under the condition (\ref{42}). 
[As seen from the arguments above, the coefficients of 
$\Omega^{-1/3}$ and $\sqrt{z}$ in Eq. (\ref{42}) 
are not exact, but they are of the order of, or approximately, 
the values in Eq. (\ref{42}).] 

The effective strength of a pointlike scatterer 
can be defined in two dimension in a similar manner. 
However, an energy-dependent correction is needed in this dimension. 
One possible way to show this is to follow the arguments 
in the previous and present sections. 
Ref.\cite{SC96} has taken this path.
Instead, we here take an alternative manner which makes it 
easy to understand the origin of the energy dependence 
specific to two dimension. 
We begin by reexamining the condition for 
the strong coupling in three dimension, Eq. (\ref{16}), 
in terms of the $\delta$-potential with a truncated basis. 
We start by rewriting Eq. (\ref{16}) as 
\begin{eqnarray}
\label{43}
\left| \bar{v}^{-1} - \bar{g}(z) \right| \alt \frac{\Delta(z)}{2},  
\end{eqnarray}
where $\bar{g}(z)$ is defined in Eq. (\ref{7}). 
Recall that it behaves like an interpolation of the inflection points of 
$\bar{G}(z)$ in Eq. (\ref{5}). 
The energy dependence of $\bar{g}(z)$ is expected to be small.  
Indeed, we have 
$\bar{g}(z) \simeq -M^{3/2}\Lambda^{1/2}/2\pi$ from Eq. (\ref{9}), 
irrespective to the energy $z$.  
This indicates that Eq. (\ref{43}) is equivalent to 
the condition (\ref{16}). 
Inserting Eqs. (\ref{7}) and (\ref{26}) into the condition (\ref{43}),  
we obtain 
\begin{eqnarray}
\label{44}
\left| v^{-1} -
\langle \varphi_{n}(\vec{x}_{0})^{2} \rangle 
P  \int_{0}^{E_{N(\Omega)}} 
\frac{\rho_{av}(E)}{z-E} dE \right| 
\alt 
\frac{\Delta(z)}{2}. 
\end{eqnarray}
This is the condition for the eigenvalue equation 
Eq. (\ref{24}) to have a solution $z$ around 
some inflection point of LHS in Eq. (\ref{24}). 
Using Eqs.(\ref{3}) and (\ref{27}), 
we have
\begin{eqnarray}
\label{45}
\left| v^{-1} - \frac{M^{3/2}}{2^{1/2}\pi^2}
F_1^{(3)}(z,E_{N(\Omega)}) \right|  
\alt \frac{\Delta(z)}{2}.  
\end{eqnarray} 
Eq. (\ref{45}) is identical to Eq. (\ref{37}) for $z \ll E_{N(\Omega)}$. 
Notice that $F_1^{(3)}(z,0)=0$, 
namely, the lower bound does not contribute on the principal integral. 

Let us now consider a two-dimensional analogue of the finite-range 
potential (\ref{21}); It takes a constant value $U_0$ 
on a finite-size region of area $\Omega$ and zero everywhere else. 
The bare strength $v$ is defined by $v=U_0 \Omega$ 
as in three dimension. 
Then, one can deduce an analogous formula to Eq. (\ref{44}) 
in two dimension.  
A crucial difference in two and three dimensions lies in 
the energy-dependence of the average level density. 
For the billiard with area $S$, it is given by $\rho_{av}=MS/2\pi$, 
according to the Weyl's formula. 
Since $\rho_{av}$ is independent of energy in two dimension, 
the analogue of Eq. (\ref{44}) is estimated by 
\begin{eqnarray}
\label{46}
F_1^{(2)}(z,E) & \equiv & \int \frac{dE}{z-E} \nonumber \\
& = & -\ln \frac{|z-E|}{\Lambda}, 
\end{eqnarray}
instead of Eq. (\ref{27}). 
Using $\langle \varphi_{n}(\vec{x}_{0})^{2} \rangle \simeq 1/S$ 
for a generic position of the scatterer, 
we obtain 
\begin{eqnarray}
\label{47}
\left| v^{-1} - \frac{M}{2\pi}\left(
F_1^{(2)}(z,E_{N(\Omega)}) -
F_1^{(2)}(z,0) 
\right) \right|  
\alt \frac{\Delta}{2}, 
\end{eqnarray} 
namely,   
\begin{eqnarray}
\label{48}
\left| v^{-1} - \frac{M}{2\pi}\left(
\ln \frac{z}{\Lambda} - \ln \frac{E_{N(\Omega)}-z}{\Lambda} \right) \right|  
\alt \frac{\Delta}{2}. 
\end{eqnarray} 
Here, the width $\Delta$ is estimated in a similar manner 
as in Eq. (\ref{15});     
\begin{eqnarray}
\label{49}
\Delta \simeq \pi^2 
\langle \varphi_{n}(\vec{x}_{0})^{2} \rangle 
\rho_{av} 
\simeq \frac{\pi M}{2}, 
\end{eqnarray}
which is independent of the energy $z$. 
The condition (\ref{48}) is identical to Eq. (51) in Ref.\cite{SC96}, 
apart from a minor change in the definition 
of the width $\Delta$ in RHS. 
In two dimension, $F_1^{(2)}(z,0)$ does not vanish and indeed 
has a logarithmic dependence on energy.  This is
the crucial difference from the two dimensional case.
At low energy $z \ll E_{N(\Omega)} \simeq 1/M\Omega$, we have 
\begin{eqnarray}
\label{50}
\left| v^{-1} - \frac{M}{2\pi}\left(
\ln \frac{z}{\Lambda} + \ln (M\Lambda\Omega) \right) \right|  
\alt \frac{\Delta}{2}. 
\end{eqnarray} 
Eq.(\ref{50}) indicates that as $\Omega$ shrinks, 
$v$ should behave like 
\begin{eqnarray}
\label{51}
v(\Omega) = 1/\left( \frac{M}{2\pi} 
\ln (M\Lambda\Omega) + r(\Omega) \right), 
\end{eqnarray}
where $r(\Omega)$ is a regular function which converges 
in the zero-size limit, $\Omega\longrightarrow 0$. 
The first term in the denominator ensures that 
the logarithmic divergence disappears 
in Eq. (\ref{50}). 
Inserting Eq. (\ref{51}) into Eq. (\ref{50}), 
we obtain in the zero-size limit, 
a two-dimensional analogue of Eq. (\ref{39}); 
\begin{eqnarray}
\label{52}
\left|  r(0) - \frac{M}{2\pi}\ln \frac{z}{\Lambda} \right|  
\alt \frac{\Delta}{2}. 
\end{eqnarray} 
This indicates that 
one can define the effective strength of the pointlike scatterer 
by 
\begin{eqnarray}
\label{53}
v_{e\!f\!f}(z) \equiv 
1/ \left( r(0) - \frac{M}{2\pi}\ln \frac{z}{\Lambda} \right).  
\end{eqnarray}
Eq. (\ref{52}) now reads  
\begin{eqnarray}
\label{54}
\left| v_{e\!f\!f}(z)^{-1} \right| \alt \frac{\Delta}{2}. 
\end{eqnarray}
Eq. (\ref{54}) with Eq. (\ref{53}) 
embodies the logarithmic strip of wave chaos that is 
the condition for the strong coupling in two dimension.  
By comparing this to Eq. (\ref{1}), we obtain 
\begin{eqnarray}
\label{55}
\bar{v}^{-1} \simeq r(0).   
\end{eqnarray}
The effective strength $v_{e\!f\!f}$ can be regarded as 
the direct measure of coupling strength of the scatterer, 
as in three dimension, 
and its inverse, $v_{e\!f\!f}^{-1}$, is a `distance' to 
a logarithmic curve of the strong coupling limit. 
The logarithmic energy-dependence in $v_{e\!f\!f}$, 
exhibits a specific feature in two dimension. 
It comes from non-vanishing $F_1^{(2)}(z,0)$ which 
can be regarded as the origin of scale anomaly 
in a formalistic sense.  Eq. (\ref{51}) shows 
\begin{eqnarray}
\label{56}
U_0 (\Omega) & = & 
1/\left( \frac{M\Omega}{2\pi} \ln (M\Lambda\Omega) + 
r(\Omega) \Omega  \right) \nonumber \\
& = & 
1/\left( \frac{M\Omega}{2\pi} \ln (M\Lambda\Omega) + 
r(0) \Omega + o(\Omega) \right). 
\end{eqnarray}
Hence, we obtain   
\begin{eqnarray}
\label{57}
\left\{ 
\begin{array}{lll}
v(\Omega)    & \longrightarrow & - 0, \\
U_0 (\Omega) & \longrightarrow & - \infty,  
\end{array}
\right.
\quad \mbox{as $\Omega \longrightarrow 0$.}  
\end{eqnarray}
This is consistent with the fact that 
a pointlike scatterer supports a single eigenstate 
with an eigenvalue smaller than $E_1$, irrespective to 
the value of formal strength $\bar{v}$ ($\neq 0$). 
The condition for the strong coupling 
in two dimension is rewritten in a comparable form with experiment. 
Inserting $v=U_0\Omega$ as well as Eq.(\ref{49}) into Eq. (\ref{50}), 
we obtain 
\begin{eqnarray}
\label{58}
\left| \frac{1}{U_0\Omega} - \frac{M}{2\pi}
\ln ( zM\Omega) \right| 
\alt \frac{\pi M }{4}   
\end{eqnarray} 
for $z\ll 1/M\Omega$. 
An arbitrary mass scale $\Lambda$ disappears from Eq. (\ref{58}). 
Similarly to the three dimensional case, 
the effects of a finite-size scatterer at low energy $z\ll 1/M\Omega$  
appear most strongly when it is weakly attractive \cite{SC96}. 

Our treatment is also applicable to one-dimensional case.
We end this section by showing that all the standard results in
the elementary textbooks on quantum mechanics for one dimensional
$\delta$-function is recovered in our formalism.
In one dimension, one can define a $\delta$-potential 
(of strength $v$) 
with the full unperturbed basis. The summation on LHS 
in Eq. (\ref{24}) is convergent in the limit of 
$N(\Omega)\longrightarrow\infty$, 
since the average level density is inversely proportional 
to square-root of energy; 
\begin{eqnarray}
\label{59}
\rho_{av}(z) = \frac{M^{1/2}L}{2^{1/2}\pi}\frac{1}{\sqrt{z}} 
\end{eqnarray}
for one-dimensional billiards with width $L$.    
The condition for the strong coupling is given by an equation formally  
identical to Eq. (\ref{44}); 
\begin{eqnarray}
\label{60}
\left| v^{-1} - 
\langle \varphi_{n}(x_0)^{2} \rangle 
P  \int_{0}^{\infty} 
\frac{\rho_{av}(E)}{z-E} dE \right| 
\alt 
\frac{\Delta(z)}{2},  
\end{eqnarray}
where $\langle \varphi_{n}(\vec{x}_{0})^{2} \rangle \simeq 1/L$ and 
the width is given by 
\begin{eqnarray}
\label{61}
\Delta(z) \simeq \pi^2 
\langle \varphi_{n}(\vec{x}_{0})^{2} \rangle 
\rho_{av} 
\simeq
\frac{\pi M^{1/2}}{2^{1/2}}\frac{1}{\sqrt{z}}. 
\end{eqnarray}
The principal integral in Eq. (\ref{60}) can 
be estimated with the use of  
\begin{eqnarray}
\label{62}
F_1^{(1)}(z,E) & \equiv & \int \frac{1}{(z-E)\sqrt{E}} dE \nonumber \\
& = & 
\frac{1}{\sqrt{z}}
\ln \left| \frac{\sqrt{z}+\sqrt{E}}{\sqrt{z}-\sqrt{E}} \right|, 
\ \ \ \ (z>0).   
\end{eqnarray}
Since we have $F_1^{(1)}(z,0) = F_1^{(1)}(z,\infty) = 0$, we get 
\begin{eqnarray}
\label{63}
\left| v^{-1} \right| \alt \frac{\Delta(z)}{2}. 
\end{eqnarray} 
Therefore, in one dimension,
the strong coupling with a pointlike scatterer
is attained when the bare strength $v$ is large.
The property is energy-independent (no scale anomaly).
Since the width becomes narrow as the energy increases,
the effect of a pointlike scatterer with any (finite) bare strength 
disappears in the high energy limit. 
The bare strength $v$ is identical to the effective 
strength $v_{e\!f\!f}$ in one dimension. 
They are related to the formal strength by
\begin{eqnarray}
\label{64} 
v^{-1} = v_{e\!f\!f}^{-1} =  \bar{v}^{-1} - 
\sum_{n=1}^{\infty} \varphi_{n}(\vec{x}_{0})^{2} 
\frac{E_{n}}{E_{n}^{2}+\Lambda^2}. 
\end{eqnarray}
In contrast to two and three dimensions, 
no divergent quantity appears in the definition
of effective coupling.  In analogy to the
similar situation in quantum field theories, one
might call this property of one-dimensional
pointlike
scatterer as {\em super-renormalizability}.  
A pointlike scatterer of bare strength $v$ 
is obtained as the zero-size limit of a finite-range $(\Omega)$ 
potential with height $U_0 \equiv v/\Omega$ in a natural manner. 
In order to ensure $v \neq 0$, $U_0$ should behaves like 
\begin{eqnarray}
\label{65} 
U_0 (\Omega) = 1 / \left(r(\Omega) \Omega \right), 
\end{eqnarray}
where $r(\Omega)$ is regular in the zero-size limit. 
Since no singular term appears in 
$v(\Omega)^{-1}$ at $\Omega \rightarrow 0$ limit, 
the usual zero-size limit,  
in which the product $U_0 \Omega$ is kept constant, 
is attained by keeping $r(\Omega)$ constant as $\Omega$ varies. 
Thus, one obtains a pointlike object 
with the bare strength 
\begin{eqnarray}
\label{66} 
v = 1 / r(0).   
\end{eqnarray}
We may conclude from the current 
perspective that it is an accidental fortune of
super-renormalizability, that has enabled the simple formulation 
of the one-dimensional Dirac $\delta$-function
with a straightforward limiting procedure.

%
%
\section{Numerical Example}

We have revealed, in Sec. II, the condition for the 
appearance of the effects of a pointlike scatterer
in three-dimensional quantum billiards. 
It has been applied to the low-energy spectrum in 
case of a small but finite-size scatterer in Sec. III. 
In this section, the predictions are confirmed by 
examining statistical properties of quantum 
spectrum. 
We restrict ourselves to the exactly pointlike case. 
Even in this case, the numerical burden of handling
very large number of basis states is quite heavy, and
a smart trick is required to overcome it.

We take a rectangular box as an outer billiard. 
We also assume the Dirichlet boundary condition 
such that wave functions vanish on the boundary. 
The mass scale is set to $\Lambda=1$ in the following. 
Fixing the value of $\Lambda$ makes all parameters dimensionless. 
The eigenvalues $E_{n}$ and eigenfunctions $\varphi_{n}(\vec{x})$  
in Eq. (\ref{4}) are given by rearranging the triple-indexed 
eigenvalues and eigenfunctions in ascending order of energy; 
\begin{mathletters}
\begin{eqnarray}
\label{67}
E_{n_{x}n_{y}n_{z}}=\frac{\pi^{2}}{2M}
\left\{
\left( \frac{n_{x}}{l_{x}} \right) ^{2} +
\left( \frac{n_{y}}{l_{y}} \right) ^{2} +
\left( \frac{n_{z}}{l_{z}} \right) ^{2} 
\right\}, &  \\
\varphi_{n_{x}n_{y}n_{z}}(\vec{x})=\sqrt{\frac{8}{V}}
\sin \frac{n_{x}\pi x}{l_{x}}
\sin \frac{n_{y}\pi y}{l_{y}}
\sin \frac{n_{z}\pi z}{l_{z}}, & 
\\ \nonumber
\ \ \ \ (n_{x,} n_{y}, n_{z} = 1,2,3,\cdots). 
\end{eqnarray}
\end{mathletters}
The mass of a particle and the side lengths of the billiard are assumed to be
$M=1/2$ and $(l_{x},l_{y},l_{z})=(1.0471976, 1.1862737, 0.8049826)$, 
respectively. In this choice of the side lengths, 
the volume of the billiard is $V=1$. 
We calculate $\bar{G}(z)$ on the interval between $E_{m}$ and $E_{m+1}$ by 
\begin{eqnarray}
\label{68}
\bar{G}(z) & \simeq & 
\langle \varphi_{n}(\vec{x}_{0})^{2} \rangle 
\int_{E_{1}}^{E_{m-2000}} 
\left( \frac{1}{z-E}+\frac{E}{E^{2}+1} \right) \rho_{av}(E)dE \nonumber \\
& + & 
\sum_{n=m-2000}^{m+2000} \varphi_{n}(\vec{x}_{0})^{2} 
\left( \frac{1}{z-E_{n}}+\frac{E_{n}}{E_{n}^{2}+1} \right) \nonumber \\ 
& + &
\langle \varphi_{n}(\vec{x}_{0})^{2} \rangle 
\\ \nonumber
&\times &
\int_{E_{m+2000}}^{\infty} 
\left( \frac{1}{z-E}+\frac{E}{E^{2}+1} \right) \rho_{av}(E)dE. 
\end{eqnarray}
When $m<2000$, the first integral is discarded and the lower bound 
of the summation is replaced by $n=1$ in Eq. (\ref{68}). 
The integral in Eq. (\ref{68}) is easily calculated by using Eq. (\ref{8}).
The approximation by Eq. (\ref{68}) serves to lessen numerical burden 
considerably, keeping a sufficient numerical accuracy. 

\begin{figure}
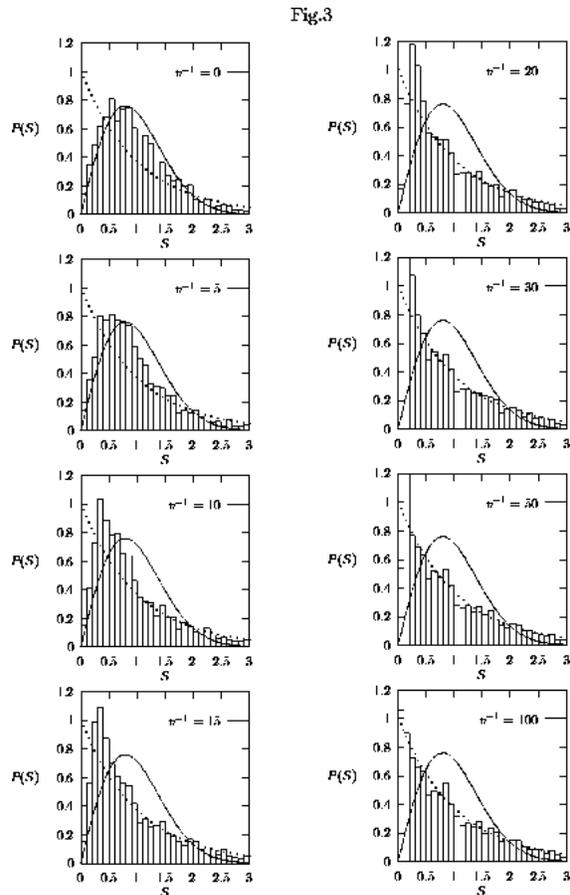

\center\psbox[hscale=0.5,vscale=0.5]{dmfig3.epsf}
\caption{
The nearest-neighbor level spacing distribution $P(S)$ is 
shown for various values of $\bar{v}^{-1}$ in case of 
the scatterer being located at the center of the rectangular solid. 
The statistics are taken within the eigenvalues between 
$z_{100}=1307.95$ and $z_{3100}=12932.70$. 
(The eigenvalues are numbered by taking into account 
only the eigenstates with even parity in each direction.) 
The solid (dotted) line is the Wigner (Poisson) distribution. 
}
\label{fig3}
\end{figure}

\begin{figure}
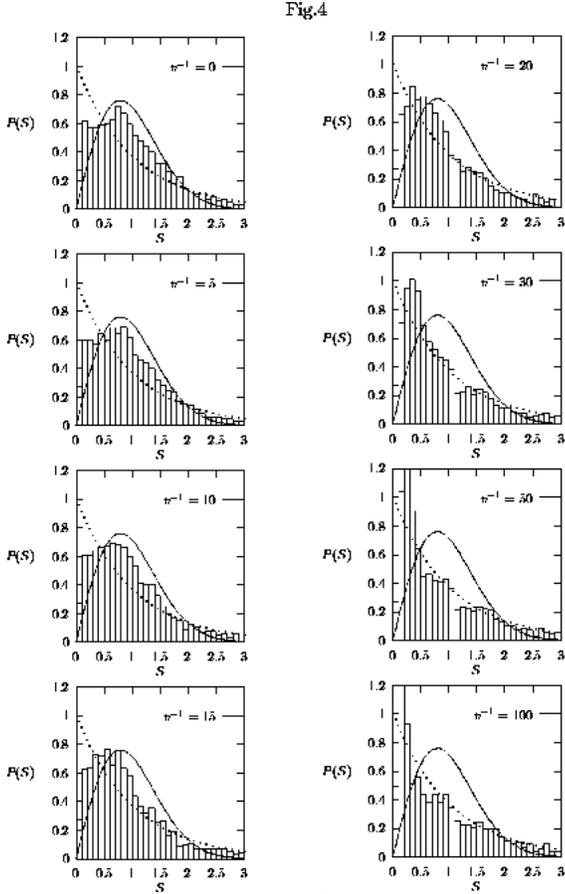

\center\psbox[hscale=0.5,vscale=0.5]{dmfig4.epsf}
\caption{
Same as Fig. 3 except for the energy 
region between $z_{17000}=40184.77$ and $z_{20000}=44767.02$. 
}
\label{fig4}
\end{figure}

For a moment, we restrict ourselves to the case where 
the scatterer is placed at the center of the billiard. 
In this case, $\langle \varphi_{n}(\vec{x}_{0})^{2} \rangle = 
\varphi_{n}(\vec{x}_{0})^{2} = 8/V$, which is eight times 
larger than the average value for generic cases. 
However, Eq. (\ref{16}) is still valid, since only eighth of
the whole unperturbed states, namely, that with 
even parity in each direction are affected by the scatterer, 
$(n_{x,} n_{y}, n_{z} = 1,3,5,\cdots)$. 
The solid curves in Fig. \ref{fig2} represent 
$\bar{v}^{-1}=-M^{3/2}/2\pi \pm \Delta (z)/2$.  
According to the condition (\ref{16}) (with $\Lambda=1$), 
the effects of a pointlike scatterer are expected to appear 
mainly in the eigenstates in the region between both curves. 
This is in fact the case as observed in Figs. \ref{fig3} and 
\ref{fig4}, where  
the nearest-neighbor level spacing distribution 
$P(S)$ is displayed for various non-negative values of $\bar{v}^{-1}$ 
in two energy regions; $z_{100}\sim z_{3100}$ in Fig. \ref{fig3} and 
$z_{17000}\sim z_{20000}$ in Fig. \ref{fig4}, respectively. 
We have numerically confirmed that 
the sign reversion of $\bar{v}^{-1}$ does not 
change the qualitative behavior of the distribution 
in both energy regions. 
Figs. \ref{fig3} and \ref{fig4} show that 
the case of $\bar{v}^{-1}=-M^{3/2}/2\pi = 
-0.056269769 \simeq 0$ is closest to the Wigner distribution 
(solid line). 
It is numerically observed that 
the second moment of $P(S)$ is given by 
$\int_{0}^{\infty} P(S) S^{2} dS \simeq 1.5$ for $\bar{v}^{-1}=0$, 
irrespective to the energy. 
This indicates that $P(S)$ is Wigner-like 
in the whole energy region for $v^{-1} \simeq -M^{3/2}/2\pi$. 
As $\bar{v}^{-1}$ increases, $P(S)$ tends to approach the Poisson 
distribution (dotted line). However, its rate depends on the energy.  
While $P(S)$ becomes intermediate in shape between the Poisson 
and Wigner distributions at $\bar{v}^{-1} \simeq 10$ in Fig. \ref{fig3},  
such distribution appears at $\bar{v}^{-1} \simeq 30$ in Fig. \ref{fig4}.  
This can be easily understood from Fig. \ref{fig2}; 
The value of $-M^{\frac{3}{2}}/2\pi + \Delta (z)/2$ is 
$11.3$ at $z_{1600}=8303.96$, and 
$25.7$ at $z_{18500}=42508.80$, respectively. 
These values can be considered as the upper bound of $\bar{v}^{-1}$ 
for inducing a Wigner-like shape in $P(S)$ at each energy region. 
With $\bar{v}^{-1}$ beyond the bound, the system is not substantially 
different from the empty billiard, and as a result, $P(S)$ resembles 
to the Poisson distribution. 
\begin{figure}
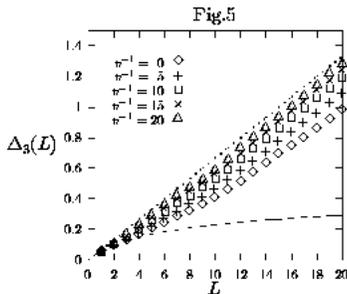

\center\psbox[hscale=0.5,vscale=0.5]{dmfig5.epsf}
\caption{
The spectral rigidity $\Delta_{3}(L)$ is 
shown for various values of $\bar{v}^{-1}$ in the energy 
region between $z_{100}=1307.95$ and  $z_{3100}=12932.70$.  
The scatterer is located at the center of the billiard. 
The solid (dotted) line is the prediction of random-matrix (Poisson) 
statistics. 
}
\label{fig5}
\end{figure}
In Fig. \ref{fig5}, the spectral rigidity $\Delta_{3}(L)$ is shown 
for various values of $\bar{v}^{-1}$. The average is taken 
in the same energy region as in Fig. \ref{fig3}. 
We can see the gradual shift to Poisson statistics (dotted line) as 
$\bar{v}^{-1}$ increases. 
Beyond $\bar{v}^{-1} \simeq 20$, the value of $\Delta_{3}(L)$ 
is close to the Poisson prediction, $L/15$. 
There still exists an appreciable difference from random-matrix 
prediction (solid line) even for the strong-coupling limit 
$(\bar{v}^{-1}=-M^{3/2}/2\pi \simeq 0)$. 
A similar tendency has been reported in two-dimensional cases \cite{S94}. 
This can be understood from the fact that 
the range of the $n$-th perturbed eigenvalue is restricted to the 
region between $n$-th and $(n+1)$-th unperturbed ones in case of 
a single pointlike scatterer. 
As a result, the number of perturbed eigenstates on a certain 
energy interval does not differ largely from the number of 
unperturbed ones in the same region. 
This restriction does not apply to the case of multiple number of
pointlike scatterers.
We can therefore expect that the increase of the number of 
scatterers makes the energy spectrum more rigid. 
For two-dimensional rectangular billiard, a recent calculation 
corroborate this argument \cite{CS96b}.

\begin{figure}
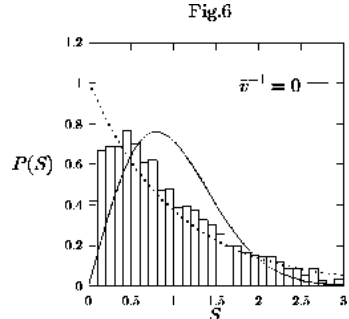

\center\psbox[hscale=0.5,vscale=0.5]{dmfig6.epsf}
\caption{
The nearest-neighbor level spacing distribution $P(S)$ is 
shown for $\bar{v}^{-1}=0$ in case of 
the scatterer being located at a generic position 
in the rectangular solid. 
The statistics are taken within the eigenvalues between 
$z_{100}=415.81$ and $z_{3100}=3503.68$. 
The solid (dotted) line is the Wigner (Poisson) distribution. 
}
\label{fig6}
\end{figure}
Up to now, we have placed a pointlike scatterer at a specific position,  
namely, the center of the rectangular box. 
We next show the level statistics for the case of a generic
location for the pointlike scatterer.
In Fig. \ref{fig6}, we show the nearest-neighbor level spacing 
distribution $P(S)$ for a box with a scatterer located at 
$\vec{x}_{0}=(0.5129731,0.5489658,0.3342914)$. 
The formal coupling is chosen to be $\bar{v}^{-1} = 0$. 
Although a nearly maximal coupling is expected to be attained  
with this value of $\bar{v}^{-1}$, 
the level repulsion is rather weak and the observed $P(S)$ 
is considerably different from the Wigner distribution. 
This can be understood by considering the state dependence 
of $\varphi_n (\vec{x}_0)^2$. 
In case that the scatterer is placed at the center, 
the value of $\varphi_n (\vec{x}_0)^2$ is 
independent of the unperturbed states; $\varphi_n (\vec{x}_0)^2 = 
8/V$ for even parity states in each direction. 
This ensures a smooth change of the value of $\bar{G}(z)$ 
at the successive inflection points. 
For a generic position of the scatterer, however, 
the value of $\varphi_n (\vec{x}_0)^2$ changes nearly at random  
as $n$ varies, causing a considerable fluctuation of 
the inflection points of $\bar{G}(z)$. As a result, 
it frequently happens that successive unperturbed states 
are not substantially affected by the scatterer 
even with the strong coupling value of the formal strength. 
It should be also noticed that, for the generic position of the scatterer, 
the width of strong coupling is substantially 
smaller than its average estimate given in Eq. (\ref{16}). 
This can be understood as follows. 
Define the width for the $n$-th state by 
\begin{equation}
\label{69}
\Delta_n (z) \equiv \pi^2 \varphi_n (\vec{x}_0)^2 \rho_{av}(z)  
\end{equation}
with $z \simeq z_n \simeq E_n$. Since $\varphi_n (\vec{x}_0)^2$ 
ranges from $0$ to $8/V$ as $n$ varies, 
the width $\Delta_n (z)$ fluctuates between $0$ and $8 \Delta (z)$ 
for a generic $\vec{x}_0$. 
Since its average is given by $\Delta (z)$, successive 
appearance of the width should be smaller than $\Delta(z)$. 
This also explains why the coupling of the pointlike scatterer 
is rather weak for the generic case.  
[For the case that the scatterer is located at the center, 
we have $\Delta_n (z) = \Delta (z)$, irrespective to the 
unperturbed states.] 
Clearly, a successive existence of the eigenstates unaffected by 
the scatterer is specific feature of a single-scatterer case.  
As the number of scatterers increases, 
such tendency disappears because only in very rare occurrence, 
none of the scatterers has a substantial influence on successive 
unperturbed eigenstates, 
as long as the coupling strength of each scatterer satisfies 
the condition (\ref{16}). 
Again, for two-dimensional cases, Numerical results  support 
this assertion \cite{CS96b}. 

%
%
\section{Conclusion}

To conclude the paper, we summarize the findings in the previous
sections.
Eq. (\ref{16}) in Sec. II is precisely the necessary condition 
for the appearance of wave chaos for three-dimensional 
pseudointegrable billiards with pointlike scatterers. 
The condition is essentially different from that 
for two dimension.
Whereas it is described by a logarithmically energy-dependent 
strip with an energy-independent width in two dimension, 
it is characterized by a parabola with a symmetric axis 
parallel to the energy axis in three dimension. 
This implies that in three-dimensional billiards, the effect of 
the pointlike scatterer is stronger in the higher energy region.
The numerical experiments using the rectangular box confirm the
assertion that even a single pointlike scatterer brings about 
wave chaos under the predicted condition,
although the precise amount of the effect depends on the location
of the scatterer. 

Since the condition for wave chaos, Eq. (\ref{16}) is described 
in terms of the formal strength $\bar{v}$ of the pointlike 
scatterer, it is not directly applicable 
to the case for realistic finite-size impurities. 
For this in mind, 
we have examined a relation between formal strength ${\bar v}$ 
of the pointlike scatterer and the bare strength $v$ 
of the finite-size potential which is defined in a natural way   
as the product of height and volume of 
the constant potential on a finite-size region.
The relation between $\bar{v}$ and $v$ also makes it clear 
how one should take a zero-size limit to obtain a pointlike object
with a given formal strength.
It is shown that $v^{-1}$ has a inverse cubic-root 
divergence in $\Omega \rightarrow 0$ limit in three dimension.
It is also shown that one can use a regular part of $v^{-1}$ 
[$r(\Omega)$ in Eq. (\ref{32})] 
as a direct measure of the coupling strength of a small scatterer.  
In other word, the inverse of the regular part corresponds to 
the effective strength of the scatterer. 
Since the coefficient of the singular part of 
$v^{-1}$ is negative, wave chaos is expected to appear at low energy 
in case of weak attraction. 

We have reviewed the previously known results in two dimension 
from the present perspective. 
Similarly to the three dimensional case,
the inverse of the bare coupling $v^{-1}$ has to contain a
singularity as a function of the size of the scatterer, and
the regular part of $v^{-1}$ plays a central role 
in determining the effective coupling strength. 
There is a crucial difference, however. 
In two dimension, a logarithmically energy-dependent correction 
term is required to define the effective strength.
The existence of the energy-dependent term results in a 
peculiar feature for two-dimension, namely, the scale anomaly. 
Its origin is identified as the $z$-dependence 
of $F_1^{(2)}(z,E=0)$ in Eq. (\ref{46}) for two dimension. There is no
corresponding term for three (and one) dimension. 

In a sense, the current work amounts to the search of a sensible
zero-size limit of small obstacles in the quantum mechanics
of general spacial dimension.  Apart from the case of one dimension,
where super-renormalizability guarantees the
existence of trivial limit ($\delta$ function), one encounters a
subtle balance of divergence and renormalizability, which results
in non-trivial properties of coupling strengths.
We hope that we have persuaded the readers
that the model of the billiards with pointlike scatterer is a valid,
mathematically sound, and practically useful idealization 
of the quantum system with small impurities.
We also hope that the predictions in this paper are to be checked
through the experiments in the laboratories. 
In particular, Eqs. (\ref{42}) and (\ref{58}) for three and two 
dimensions respectively, can be directly tested, since they are stated 
in an experimentally controllable form. 
Recent progress of microwave techniques with macroscopic devices 
\cite{SS90,S91,Getal92} 
offers a possible opportunity.  

\vspace*{5mm}

Numerical computations have been performed on the HITAC MP5800 and S-3800 
computers at the Computer Centre, the University of Tokyo. 
We thank Prof. Izumi Tsutsui for helpful discussions.

%
%

\begin{references}
\bibitem{NT00}
The readers are warned for 
a confusion in the appellation of the
coupling strengths in our previous paper.
The formal strength ${\bar v}$ is refered to 
as ``bare strength $v_B$'' in Ref. \cite{S94}.
There, the bare strength $v$ in current terminology is
called ``physical strength $v_P$''.
\end{references}
\end{document}